\PassOptionsToPackage{hyphens}{url}
\documentclass[aps,rmp,amsfonts,amssymb,amsmath,float,twocolumn,showkeys,longbibliography
]{revtex4-1}
\ifx\pdftexversion\undefined
\usepackage[dvips]{graphicx}
\else
\usepackage[pdftex]{graphicx}
\fi

\usepackage{graphicx}
\begin{document}
\title{Role of a Habitat's Air Humidity in Covid-19 Mortality.}

\author{Irina V. Biktasheva$^{1,2,}$}
\email{ivb@liverpool.ac.uk}
\affiliation{$^{1}$Department of Computer Science, University of Liverpool, Liverpool L69 3BX, UK}
\affiliation{$^{2}$CEMPS, University of Exeter, Exeter EX4 4QF, UK}
\date{\today}

\begin{abstract}
Transient local over-dry environment might be a contributor and an explanation for the observed asynchronous local rises in Covid-19
mortality. We propose that a habitat's air humidity negatively correlate
with Covid-19 morbidity and mortality, and support this hypothesis on the example
of publicly available data from German federal states. 
\end{abstract}

\keywords{COVID-19 Mortality; Habitat; Air Humidity; Negative Correlation}

\maketitle

% \newpage

\paragraph*{Introduction.} Covid-19 virus~\cite{C-19_Nature} is transmitted through droplets which last longer in
humid air.
Therefore, humidity is believed to be pro-Covid-19
infection and mortality.
There are, however, data that contradict this belief. 
For instance, Wuhan, where
Covid-19 was first identified and studied, is in humid subtropical
climate zone~\cite{ClimateWuhan}, but December, when mortality sharply
raised, is the driest month of the year there. The purpose of this communication is to present and
substantiate a viewpoint that air humidity negatively correlate with Covid-19 morbidity and mortality. 

\paragraph*{The main hypothesis} consists of two parts, of different degree of plausibility. 
First, mucous membranes of the upper respiratory tract present the
  first and essential barrier against Covid-19 virus entering
  human organism. Hence the state of the mucous membranes is a correlate to organism's resistivity. 
Second, a dry season normally cause respiratory mucosa to become over-dry. In
  presence of Covid-19 virus the latter might become a factor of massive fatality.

\paragraph*{Direct evidence in support of the hypothesis.}
\begin{figure}[tb]
  \centering
  \includegraphics{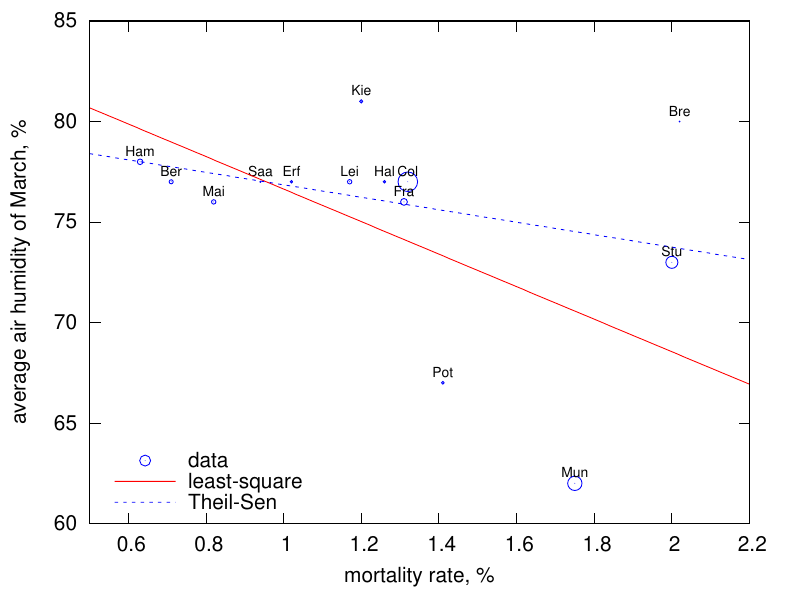}
  \caption[]{Air humidity in March vs Covid-19 mortality in German
    federal states (circles). The size of
    the circles is proportional to the population of the federal
    state; the labels are abbreviations for the largest cities of the
    federal lands. Red solid line shows the linear
    regression. Theil-Sen regression is shown with the blue dashed line. 
  }
  \label{mort-hum}
\end{figure}
Correlation of Covid-19 with age, both in terms of registered
  cases and of mortality is the first and best known fact about this
  strain~\cite{C19_AgeFactor}. The state of respiratory mucosa also correlates with age~\cite{Beule_2010}.

  Correlation of Covid-19 mortality with low air
  humidity is less obvious.
By way of anecdotal evidence, in addition to the coincidence of the
  beginning of the epidemic with the dry season in Wuhan mentioned
  above, note that Seoul and especially Tokyo, where the
  incidence and mortality have been lower, have on average wetter
  climate~\cite{ClimateSeoul, ClimateTokyo}. 
  Correlate of sharp raise in Covid-19 mortality with local dry period
  may also be seen on the example of Lombardy, where February is the
driest month in Milan~\cite{ClimateMilan}, as opposed to a wetter
beginning of the spring in Rome~\cite{ClimateRome}. The much dryer March in Spain as opposed to
e.g. more humid neighbouring Portugal~\cite{ClimateLisbon} seem to point in the
same direction of raise in Covid-19 mortality correlate with transient over-dry
local environment.

As an illustrative and preliminary example of a more systematic evidence, we have considered
Covid-19 mortality rate, defined as the number of deaths per number of
confirmed infections, in German federal
states~\cite{wikipedia-pandemic-germany} where the majority of deaths
happened last March 2020.  FIG.\ref{mort-hum} shows the Covid-19 mortality rate
in the federal states vs the average local air humidity in March. The
local air humidity was defined as proxy recorded in the
largest city of each federal state~\cite{weather-and-climate}. The
choice of the German federal
states data sets is motivated by the data availability and
reliability, and, in particular, by the presumed uniformity of the data collection
protocols in Germany. Mecklenburg-Vorpommern has been excluded as the resource
\cite{weather-and-climate} gives no humidity data neither for Rostock
nor for Schwerin. The linear least squares fit (red solid line), weighted by the most recent population size
of the federal states as given by Wikipedia, gives the slope of
$-8.09$ with a standard deviation of $\pm3.32$, i.e. reliably
negative. We have also applied the Theil-Sen estimator, also weighing
the data points proportionally to population sizes, which gives the
slope of $-3.10$. The corresponding fits are also shown in
FIG.\ref{mort-hum}  (blue dashed line). The discrepancy between the linear and Theil-Sen
estimates are not surprising as the problem is clearly multi-factorial
and we are looking at only one of the factors. Still, the two
estimates concur that the slope of the dependence is negative, that is
mortality is on average higher in a drier air and lowers with rise of air humidity,
which confirms our hypothesis on negative correlate of Covid-19
mortality with local air humidity.

Note that~\cite{Klein-etal-2020} presents evidence of negative
correlation of air humidity with Covid-19 transmission
rate based on all-China data. A similar study \cite{SciTotEnv_C-19_Mar2020} based on Wuhan
data indicates negative correlation of air humidity with Covid-19
mortality. Both of the studies did not seem to take into
account the effect of non-causal correlation between the seasonal increase of humidity and decrease in
transmission rate due to the taken administrative measures. Still,
both studies are interesting as they present an approach complementary
to the one we used in  FIG.\ref{mort-hum}, namely, both studies do not take into account regional
variations of air humidity and Covid-19 statistics; instead,
\cite{SciTotEnv_C-19_Mar2020} is exploiting their temporal
variations.
In any case, the overall conclusion from those studies
concurs with our hypothesis.

The above is ``direct'' evidence as it allows one to hypothesise direct causal
relationship: dry air causing over-dry respiratory mucosa in older and vulnerable population causing increase of Covid-19 mortality.

\paragraph*{Indirect evidence in support of the hypothesis.}
Dry nasal mucosa correlates with loss of smell and
  taste~\cite{Beule_2010}, and loss of smell and
  taste correlates with Covid-19
  statistics~\cite{C19_SmellSymptom}. 

  Dry air is a known
  risk factor for dry eyes~\cite{DryAir_DryEyes}, with dry eyes being a risk factor for
  conjunctivitis~\cite{DryEyes_conjuctivitis}. Recent reports show correlates of Covid-19 statistics with
  conjunctivitis~\cite{C19_Occular, C19_conjunctivitis}.

  Some groups of patients identified as particularly vulnerable
  to Covid-19, e.g. diabetes~\cite{C19_AgeFactor}, also correlate with the diminished
  function of respiratory mucosa~\cite{Beule_2010}.

  For these above evidence, the direction of causal links, if any, is less clear. 

\paragraph*{Verifiable predictions of the hypothesis.}

A direct verification of the proposed hypothesis would be analysis of
the instant local air humidity in the statistics of Covid-19 incidence and
mortality. A statistically significant correlation would confirm the
hypothesis.

A more sophisticated way of checking the hypothesis would be via
spatiotemporal modelling of the pandemic. Such modelling, which no
doubt will be attempted by many research groups, will be most successful if
and when it takes into account as many relevant factors as
possible. Hence if the proposed hypothesis is true, taking into
account the air humidity of the habitat would improve the quality and
predictive power of the models.

\paragraph*{Discussion and practical consequences.}

Air humidity depends on multiple parameters: the local
instant and annual rate of precipitation, diurnal and annual temperature range, 
altitude, etc. That is why it is so difficult to estimate air
humidity based on e.g. local precipitation only. 

If the negative correlate of Covid-19 mortality with air humidity is verified, it might suggest certain
practical steps in addition to the medical and administrative measures
already in place, and those yet to be
proposed based on other considerations. 

%There is yet too little spatiotemporal data to model global spread
%of Covid-19 virus and the observed asynchronous peaks of mortality. 

Iceland's Covid-19 screening
  showed people of all ages equally susceptible for the
infection~\cite{C-19_Iceland}, with about $50\%$ of those infected having no symptoms at the time of
testing~\cite{C-19_Iceland_50noSymptoms}. In the view of the latter, the world wide correlation of Covid-19 with age
might appear to be skewed towards those patients showing more symptoms and
therefore more tested.
Therefore, it might be useful to distinguish between the spread of
Covid-19 infection and the asynchronous local rises of Covid-19
mortality. For instance, a prolonged dry weather may be taken as an indication
of likely local elevation of Covid-19 mortality. In Madrid, August is the
driest month of the year~\cite{ClimateMadrid}, so preventative
measures might be indicated to forestall or 
flatten the second wave of Covid-19 there. On a general point, it might take at least an annual cycle of the global data to fully appreciate
the spatiotemporal pattern of Covid-19 pandemic, and build the data
based model. 

In presence of Covid-19 virus, patients with tendency to dry respiratory mucosa might be particularly
vulnerable to the exposure to dry air. Indoor environment might become
over-dry due to the winter central heating, domestic devices producing heat, especially if the only source
of the indoor humidity are people themselves which might be not
enough. It might be not possible to alter local microclimate, not to mention
an instant change of the global one. However, control of indoor environment is feasible and might mitigate patients'
exposure to Covid-19. Balance of exposure to Covid-19 virus in dry air against the well known exposure
to bacterial infection in a humid environment must be taken into
account when developing a healthy indoor technology. 
To the author's knowledge, this aspect has not yet been discussed, while, if true, this could immediately start saving lives, which is the reason for this publication.

\paragraph*{Availability of supporting data.}
The datasets used and/or analysed in this paper are available from the websites.

\paragraph*{Declaration of interests.}
The author declares that she has no known competing financial interests or personal relationships that could have appeared to influence the work reported in this paper.

\paragraph*{Acknowledgement}
This paper is entirely the author's initiative. It does not result from
any funded research project. The author is grateful to Professor
Vadim Biktashev, Dr Sergey Blagodatsky, and Dr Evgenia Blagodatskaya, for much valued discussion of this paper.

\onecolumngrid

% \section*{References}
% \bibliographystyle{apsrmp4-1}
\bibliographystyle{aipauth4-1}
\bibliography{c_19} 

\end{document}